# Phonon Raman scattering in LaMn$_{1-x}$Co$_x$O$_3$ ($x$ = 0, 0.2, 0.3, 0.4, and 1.0)

V.P. Gnezdilov,[1] Yu.G. Pashkevich,[2] A.V. Yeremenko,[1] P. Lemmens,[3,4] G. Güntherodt,[3] S.V. Shiryaev,[5] G.L. Bychkov,[5] and S.N. Barilo[5]

[1]) B.I. Verkin Inst. for Low Temp. Physics NASU, 61164 Kharkov, Ukraine;
[2]) A.A. Galkin Donetsk Phystech NASU, 83114 Donetsk, Ukraine;
[3]) 2. Physikalisches Institut, RWTH Aachen, D-52056 Aachen, Germany;
[4]) Max Planck Institute for Solid State Research, D-70569 Stuttgart, Germany;
[5]) Institute of Physics of Solids & Semiconductors, Academy of Sciences, 220072 Minsk, Belarus

The Raman-active phonons in perovskite-like LaMn$_{1-x}$Co$_x$O$_3$ ($x$ = 0, 0.2, 0.3, 0.4, and 1.0) were studied by measuring Raman spectra at temperatures of 295 and 5 K. The changes in the spectra with Co doping are correlated with the decrease of orthorhombic distortions. Surprisingly more phonon lines than allowed for the rhombohedral LaCoO$_3$ structure were observed in the spectra.

The manganese perovskites of the type $R_{1-x}A_x$MnO$_3$ ($R$= rare earth, $A$ = Ca, Sr, Ba, or Pb) have been a subject of scientific investigations for many decades. Recently, they attached a renewed interest due to the observation of a colossal magnetoresistance,[1,2] charge, spin, and orbital ordering effects as a function of Mn$^{3+}$/Mn$^{4+}$ ratio.[3-5] Another system, Mn-site-doped, with formula LaMn$_{1-x}$D$_x$O$_3$ ($D$ = Cr, Fe, Co, or Ni) was intensively studied in the 60's, but colossal magnetoresistance was not mentioned in them until the 90's.[6] While Raman spectra of La-site-doped compounds have been reported in numerous of publications[7-13], surprisingly nothing was done on Mn-site-doped compounds. In this work we report the results of the optical phonons study in the perovskite oxides LaMn$_{1-x}$Co$_x$O$_3$ ($x$ = 0.0, 0.2, 0.3, 0.4, and 1.0). The end member of this system, namely LaCoO$_3$ have been the subject of continuing interest since the 50's due to unusual magnetic properties and two spin-state transitions discovered.[14,15]

Raman scattering measurements were carried out in quasi-backscattering geometry using 514.5 nm argon laser line. The incident laser beam of 10 mW power was focused onto a 0.1 mm spot of the mirror-like chemically etched as grown crystal surface. The sample was mounted on the holder of a He-gas-flow cryostat using silver glue. The scattering light was analyzed with a DILOR XY triple spectrometer combined with a nitrogen-cooled CCD detector. Provided the naturally grown surfaces of the perovskitelike crystals are the quasicubic ones, the measurements were done in the $xx(zz)$ scattering configuration, where $x$ and $z$ are the [100] and [001] quasicubic directions, respectively.

The structural properties of LaMn$_{1-x}$Co$_x$O$_3$ were characterized in several publications.[15-19] The crystal structures of the end members, LaMnO$_3$ and LaCoO$_3$ were found to be orthorhombic (space group *Pnma*, Z = 4)[16] and rhombohedral (space group $R\bar{3}c$, Z = 2),[17] respectively. The compounds with 0.15<$x$<0.50 had been found to be orthorhombic.[18,19] When $x$>0.50 and near to 1.0, the compounds have rhombohedral structure.[19] Around 0.5 doping level, compounds have the mixture of two structural phases, orthorhombic and rhombohedral.[19] The idealized cubic perovskite structure of the LaMn(Ca)O$_3$ crystal is shown in Fig. 1. The orthorhombic *Pnma* structure can be obtained by two consequent rotations of the Mn(Co)O$_6$ octahedra around the [010] and [101] directions of cubic perovskite. The rhombohedral $R\bar{3}c$ structure is generated by the rotation of the same octahedral about the cubic [111] direction.

Results of group-theoretical analysis for zone-center vibrations are presented in Table 1 for orthorhombic LaMnO$_3$ and rhombohedral LaCoO$_3$. Of the total 30 Γ-point phonon modes, only 5 ($A_{1g}$ + 4$E_g$) are Raman active for rhombohedral LaCoO$_3$ structure and of 60 Γ-point phonon modes, 24 (7$A_g$ + 5$B_{1g}$ + 7$B_{2g}$ + 5$B_{3g}$) are Raman-active for orthorhombic LaMnO$_3$ structure. The increase in the phonon modes from 5 to 24 on going from rhombohedral to orthorhombic structure is due to (i) lowering of crystal symmetry which splits the doubly degenerate $E_g$ modes into nondegenerate $B_{2g}$ + $B_{3g}$, (ii) displacement of oxygen atoms into the lower symmetry site of the La-plane which introduces new Raman-active vibrations, (iii) doubling of the unit cell which folds the zone-boundary modes of the rhombohedral structure into zone-center modes of the orthorhombic structure.

Raman spectra of LaMn$_{1-x}$Co$_x$O$_3$ compounds at 295 K are shown in Fig. 2. Room temperature measurements, lattice dynamical calculations, and assignment of the Raman modes of undoped LaMnO$_3$ were done previously by Iliev et al.[20] Our spectrum of LaMnO$_3$ is consistent with spectra at 300 K reported earlier.[9,20] The spectrum exhibit three broad bands centered near 280, 490, and 610 cm$^{-1}$. The line near 280 cm$^{-1}$ was assigned to a rotationlike.[20] The other two bands near 490 cm$^{-1}$ and 610 cm$^{-1}$ are related to the Jahn-Teller octahedral distortions.[20] Given that the Jahn-Teller distortions are static and ordered in orthorhombic LaMnO$_3$, these bands are Raman-allowed modes bending and stretching type, respectively. The fours peak in the spectrum at ~ 310 cm$^{-1}$ is associated with vibrations of apex oxygen (O$_1$) atoms along the $x$ direction. The spectra of the $x$ = 0.2, 0.3, and 0.4 samples look quite similar to the $x$ = 0 sample, except for the width and position of the bands in the region of 500 and 600 cm$^{-1}$. At room temperature the spectra of the LaCoO$_3$ sample exhibit peaks centered near 130, 160, 480, 555, 610, and 780 cm$^{-1}$, and four broad bands at 70, 270, 340, and 400 cm$^{-1}$.

Lowering the temperature, more phonon peaks become visible in the Raman spectra of LaMn$_{1-x}$Co$_x$O$_3$. In Fig. 3 we present spectra measured at 5 K.

*LaMnO$_3$*: The spectra LaMnO$_3$ sample shows 10 resolved peaks at 80, 110, 130, 154, 184, 257, 280, 314, 496, and 610 cm$^{-1}$. The strong high frequency lines in the studied sample are broader than the corresponding lines in the spectra of measured earlier.[9,20] The reason of this broadening is the presence of small amount of excess oxygen in our sample. Using X-ray diffraction, magnetic susceptibility, and chemical analysis, the oxygen content was estimated as 3.071.

*LaMn$_{1-x}$Co$_x$O$_3$* ($x$ =0.2, 0.3, 0.4): It is well known that Raman spectroscopy is a sensitive tool for the study of both local and spatially coherent structural changes. Spectra of the samples with $x$ > 0 differ from these one of pure LaMnO$_3$ and in the following we will concentrate on the effect of the Mn substitution by Co on the rotation-, bending-, and stretching-like vibrations of the $MnO_6$ octahedra. The rotational mode at 280 cm$^{-1}$ in $LaMnO_3$ shifts to the lower energy (~ 270 cm$^{-1}$) in the samples with $x$ = 0.2, 0.3, and 0.4. The frequency of this mode is a measure of the degree of the rotational distortions (the averaged angle of octahedral tilts). For example, comparing the spectra of more distorted $YMnO_3$ and less distorted $LaMnO_3$ a large shift from 396 cm$^{-1}$ to 284 cm$^{-1}$ was observed.[20] So, the softening of the rotation-like mode with the Co doping is the indication of the orthorhombic distortions decreasing.

The scattering intensity of the phonon modes at 496 and 610 cm$^{-1}$ decreases in the Co-doped samples. The decreasing of intensity of these modes reasonably be related with the reduction of thee Jahn-Teller distortions in the averaged structure, introduced by the presence of Co. Moreover, each of these modes splits into two components. This splitting can be attributed to the coexistence of two types of octahedra, namely $MnO_6$ and $CoO_6$ in the composition. The frequency positions of some high-frequency lines are plotted on Fig. 4.

*LaCoO$_3$*: The spectra of $LaCoO_3$ are however somewhat surprising since (i) it is not similar to the spectra of isostructural rhombohedral $LaMnO_{3+\delta}$,[9,10,21,22] $La_{1-x}A_xMnO_3$,[9,10] and $LaAlO_3$;[22] (ii) due to the neutron-diffraction data the sample should have the rhombohedral symmetry $D_{3d}^6$ and only 5 Raman-active phonon modes were expected. The main peaks in the spectra are at 75, 135, 169, 486, 560, 656, and 785 cm$^{-1}$. Note that high-frequency lines, intensive in $LaCoO_3$ even at room temperature, were not observed in the spectra of isostructural compounds. Moreover, wide bands centered approximately at 198, 281, 366, 407 cm$^{-1}$ and shoulder at 702 cm$^{-1}$ present in the spectrum. Probably part of the features observed in the spectrum are a contribution of the second-order Raman process, but it is more reasonable to assume that the $LaCoO_3$ crystal has a lower symmetry than rhombohedral one. Precise measurements of polarized Raman spectra could provide valuable information about the structure of this compound.

In conclusion, we have measured the Raman phonons in the $LaMn_{1-x}Co_xO_3$ ($x$ = 0, 0.2, 0.3, 0.4, and 1.0) system at temperatures 295 and 5 K. Characteristic changes have been observed in the phonon spectra with the Co concentration and correlated with reduction of the octahedral distortions. The spectrum of pure $LaCoO_3$ exhibit more peaks than allowed for the rhombohedral ($D_{3d}^6$) symmetry and more accurate definition of the crystal structure of this compound is necessary.

This work was supported by INTAS Grant N 01-0278 and NATO Collaborative Linkage Grant PST.CLG.977766.

TABLE 1. Factor group analysis and selection rules for the zone-center vibrational modes of the orthorhombic $LaMnO_3$ and rhombohedral $LaCoO_3$.

| Sample and space group | Atom | Number of equivalent positions (Wyckoff notation) | Site symmetry | Irreducible representation of modes | Activity and selection rules |
|---|---|---|---|---|---|
| LaMnO$_3$ *Pnma* ($D_{2h}^{16}$), Z = 4 | La | 4(c) | $C_s$ | $2A_g + B_{1g} + 2B_{2g} + B_{3g} + A_u + 2B_{1u} + B_{2u} + 2B_{3u}$ | $\Gamma_{Raman} = 7A_g + 5B_{1g} + 7B_{2g} + 5B_{3g}$ |
| | Mn | 4(b) | $C_i$ | $3A_u + 3B_{1u} + 3B_{2u} + 3B_{3u}$ | $\Gamma_{IR} = 9B_{1u} + 7B_{2u} + 9B_{3u}$ |
| | O$_1$ | 4(c) | $C_s$ | $2A_g + B_{1g} + 2B_{2g} + B_{3g} + A_u + 2B_{1u} + B_{2u} + 2B_{3u}$ | $\Gamma_{acoustic} = B_{1u} + B_{2u} + B_{3u}$ |
| | | | | | $\Gamma_{silent} = 8A_u$ |
| | O$_2$ | 8(d) | $C_1$ | $3A_g + 3B_{1g} + 3B_{2g} + 3B_{3g} + 3A_u + 3B_{1u} + 3B_{2u} + 3B_{3u}$ | $A_g$: $a_{xx}, a_{yy}, a_{zz}$ $B_{1g}$: $a_{xy}$; $B_{2g}$: $a_{xz}$; $B_{3g}$: $a_{yz}$ |
| LaCoO$_3$ $R\bar{3}c$ ($D_{3d}^6$), Z = 2 | La | 2(b) | $C_{3i}$ | $A_{1u} + A_{2u} + 2E_u$ | $\Gamma_{Raman} = A_{1g} + 4E_g$ |
| | Co | 2(a) | $D_3$ | $A_{2g} + A_{2u} + E_g + E_u$ | $\Gamma_{IR} = 3A_{2u} + 5E_u$ |
| | O | 6(e) | $C_2$ | $A_{1g} + 2A_{2g} + 3E_g + A_{1u} + 2A_{2u} + 3E_u$ | $\Gamma_{acoustic} = A_{2u} + E_u$ |
| | | | | | $\Gamma_{silent} = 3A_{2g} + 2A_{1u}$ |
| | | | | | $A_{1g}$: $a_{xx} + a_{yy}, a_{zz}$ |
| | | | | | $E_g$: $(a_{xx} - a_{yy}, a_{xy}), (a_{xz}, a_{yz})$ |

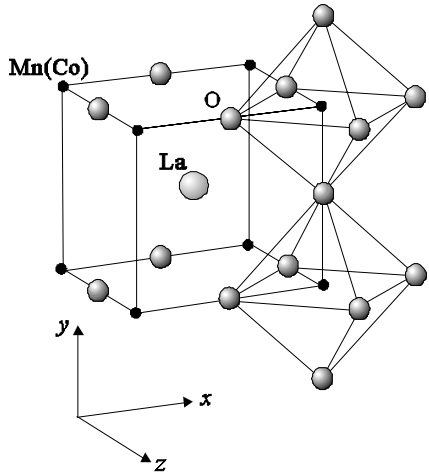

Fig. 1. The unit cell of the simple perovskite structure.

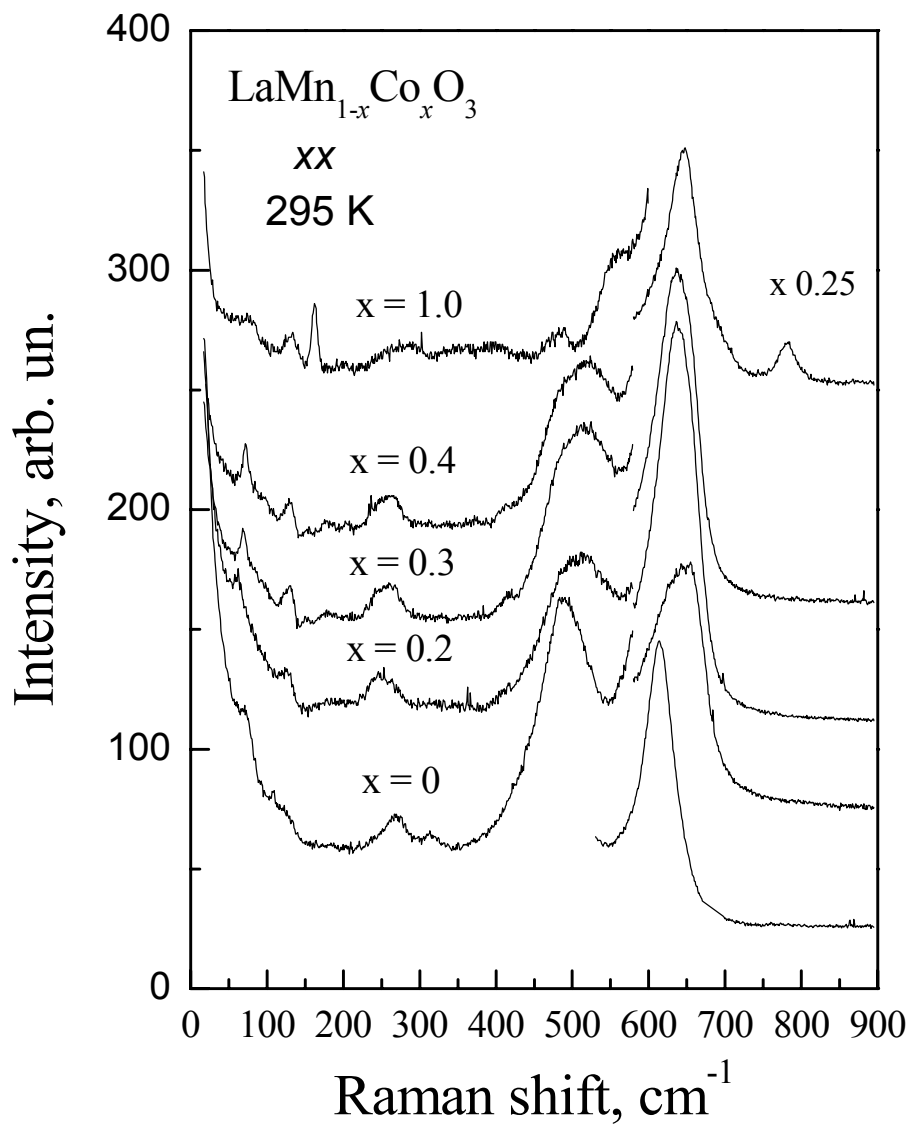

Fig. 2. Raman spectra of single crystals of LaMn$_{1-x}$Co$_x$O$_3$ at 295 K. All spectra are shifted for convenience. The right part of the spectra are multiplied by a factor indicated there.

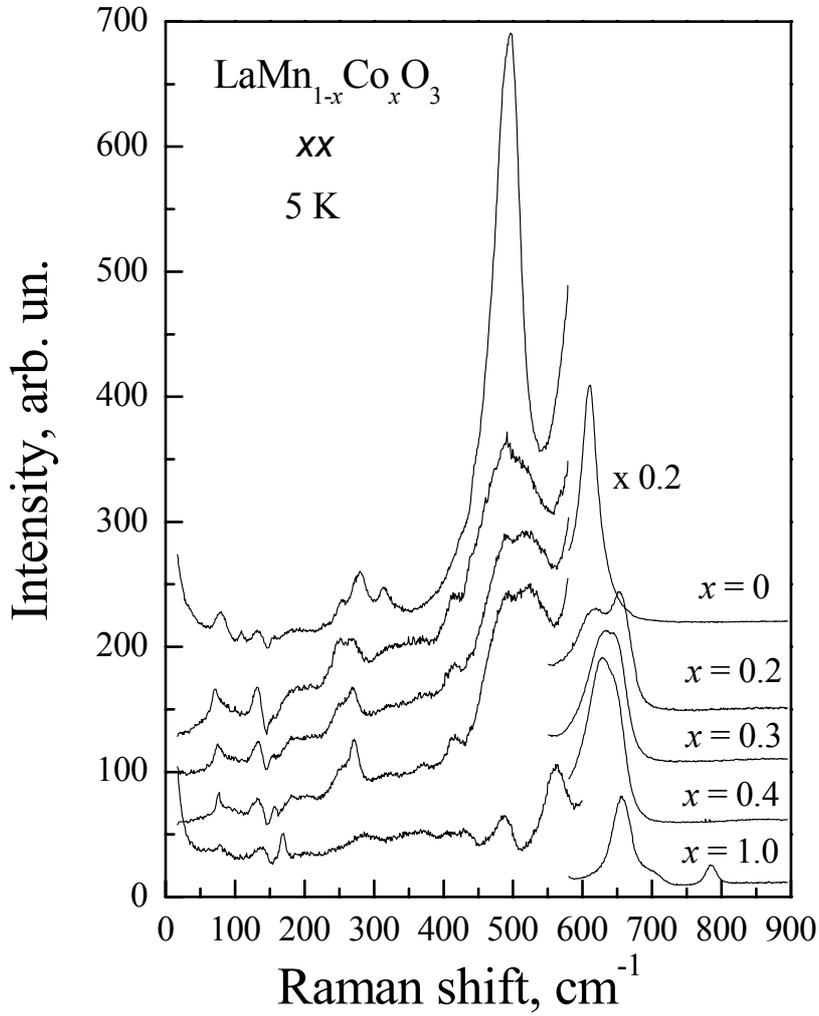

Fig. 3. Raman spectra of single crystals of LaMn$_{1-x}$Co$_x$O$_3$ at 5 K. All spectra are shifted for convenience. The right part of the spectra are multiplied by a factor indicated there.

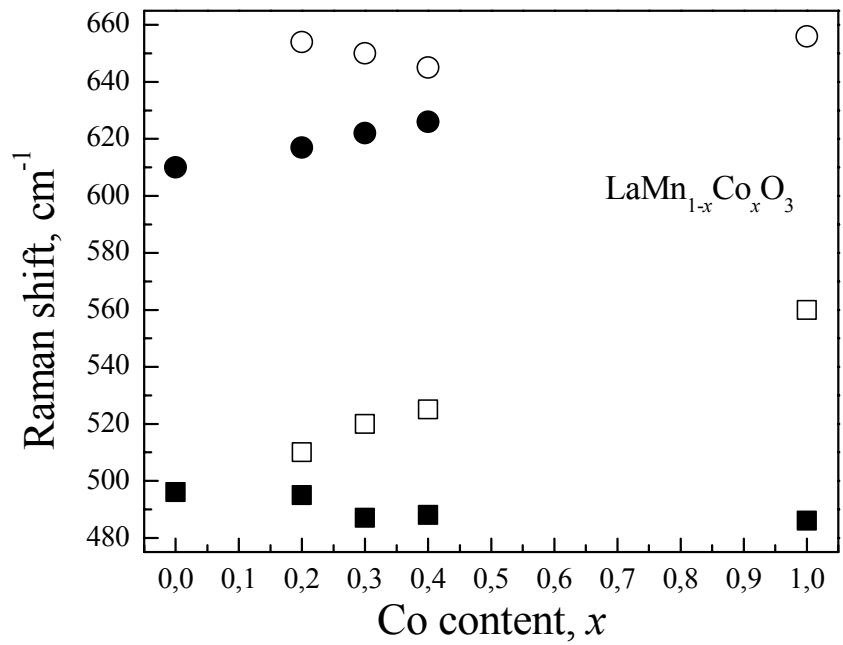

Fig. 4. Raman shift of some high-frequency phonon modes versus $x$ in the LaMn$_{1-x}$Co$_x$O$_3$ single crystals.